# Morphology of the Interstellar Cooling Lines Detected by $COBE$[1]

C. L. Bennett[2], D. J. Fixsen[3], G. Hinshaw[4], J. C. Mather[2], S. H. Moseley[2], E. L. Wright[5], R. E. Eplee, Jr.[6], J. Gales[3], T. Hewagama[4], R. B. Isaacman[6], R. A. Shafer[2], and K. Turpie[6]

astro-ph/9311032  15 Nov 1993

---

[1] The *Cosmic Background Explorer*, *COBE*, is supported by NASA's Astrophysics Division. The Goddard Space Flight Center (GSFC), under the scientific guidance of the *COBE* Science Working Group, is responsible for the development and operation of *COBE*.

[2] Code 685, Infrared Astrophysics Branch, Goddard Space Flight Center, Greenbelt, MD 20771, E-mail I:bennett@stars.gsfc.nasa.gov

[3] Applied Research Corp., Code 685.3, NASA/GSFC, Greenbelt, MD 20771

[4] Hughes-STX Corp., Code 685.9, NASA/GSFC, Greenbelt, MD 20771

[5] UCLA Astronomy Dept., Los Angeles, CA 90024-1562

[6] General Sciences Corp., Code 685.3, NASA/GSFC, Greenbelt, MD 20771

# ABSTRACT

The FIRAS instrument on the *COBE* satellite has conducted an unbiased survey of the far-infrared emission from our Galaxy. The first results of this survey were reported by Wright et al. (1991). We report the results of new analyses of this spectral survey, which includes emission lines from 158 $\mu$m $C^+$, 122 $\mu$m and 205 $\mu$m $N^+$, 370 $\mu$m and 609 $\mu$m $C^0$, and CO J=2-1 through 5-4. We report the morphological distribution along the galactic plane ($b = 0°$) of the spectral line emission, and the high galactic latitude intensities of the $C^+$ and 205 $\mu$m $N^+$ emission. In the galactic plane the 205 $\mu$m line of $N^+$ generally follows the 158 $\mu$m $C^+$ line distributions, but the intensities scale as $I(N^+\ 205\ \mu m) \propto I(C^+\ 158\ \mu m)^{1.5}$ towards the inner Galaxy. The high galactic latitude intensity of the 158 $\mu$m fine structure transition from $C^+$ is $I(C^+\ 158\ \mu m) = (1.43 \pm 0.12) \times 10^{-6} \csc|b|$ erg cm$^{-2}$ s$^{-1}$ sr$^{-1}$ for $|b| > 15°$, and it decreases more rapidly than the far infrared intensity with increasing galactic latitude. $C^+$ and neutral atomic hydrogen emission are closely correlated with a $C^+$ cooling rate of $(2.65 \pm 0.15) \times 10^{-26}$ erg s$^{-1}$ H atom$^{-1}$. We conclude that this emission arises almost entirely from the Cold Neutral Medium. The high galactic latitude intensity of the 205 $\mu$m fine structure transition from $N^+$ is $I(N^+\ 205\ \mu m) = (4 \pm 1) \times 10^{-8} \csc|b|$ erg cm$^{-2}$ s$^{-1}$ sr$^{-1}$ arising entirely from the Warm Ionized Medium. We estimate the total ionizing photon rate in the Galaxy to be $\phi = 3.5 \times 10^{53}$ ionizing photons per second, based on the 205 $\mu$m $N^+$ transition.

*Subject headings:* ISM: atoms, bubbles, general, molecules — line: identification Galaxy: general

## 1. Introduction

An unbiased far infrared survey of the spectral line emission from our Galaxy has been reported from the Cosmic Background Explorer (*COBE*) mission by Wright et al. (1991). Unlike other observations that concentrate on specific small regions of the sky at particular wavelengths, the *COBE* FIRAS instrument observed nearly the entire sky with nearly two decades of wavelength coverage. Thus, the *COBE* data show that the large scale diffuse emission from our Galaxy is dominated by a far infrared continuum that cools dust grains, and the 158 $\mu$m ground state transition of $C^+$ that cools the neutral gas in the Galaxy. Two transitions of $N^+$ trace the large-scale low-density extended ionized component of the Galaxy. In addition, the 370 $\mu$m and 609 $\mu$m lines of $C^0$ and the J=2-1 through 5-4 lines of molecular CO were detected, and trace the interiors of neutral regions.

Mechanisms for heating interstellar gas include collisions, radiation from stars, shocks, and cosmic rays. Examination of the spectral lines that cool the gas can help determine the dominant excitation mechanisms and conditions. Studies of particular bright regions in our Galaxy and observations of external galaxies have suggested that stellar ultraviolet radiation can ionize vast volumes of a galaxy and that the far ultraviolet (FUV) radiation impinging on neutral cloud surfaces is responsible for a large fraction of the observed far infrared (FIR) spectral line emission that cools the gas.

Tielens & Hollenbach (1985a) defined *photodissociation regions* (hereafter referred to as PDRs) as neutral "regions where FUV radiation dominates the heating and/or some important aspect of the chemistry. Thus photodissociation regions include most of the atomic gas in a galaxy, both in diffuse clouds and in the denser regions..." Early work on PDR physics includes Glassgold & Langer (1974, 1975, 1976), Langer (1976), Black & Dalgarno (1977), Clavel, Viala & Bel (1978), and de Jong, Dalgarno & Boland (1980). Tielens & Hollenbach (1985a) present a detailed and quantitative model for one dimensional PDRs.

The basic photodissociation process can be summarized as follows. Far ultraviolet photons in the energy range 6-13.6 eV, often from O and B stars, impinge on the surfaces of neutral clouds where they will be stopped by impacts with dust grains. The dust grains emit photoelectrons and cool by continuum far infrared emission. The UV generally photodissociates molecules and photoionizes atoms with ionization potentials less than the Lyman limit. The photoelectrons ejected from dust grains heat the gas, resulting in the excitation of $C^+$ atoms from their ground state to their first excited state, $\Delta E/k = 91$ K above the ground state. $C^+$ cools by the emission of 158 $\mu$m radiation. The entire process thus converts far UV radiation to far infrared continuum and spectral line emission.

Only far UV radiation can contribute to this process since the typical work function of an interstellar grain is ∼6 eV (de Jong 1980). The photoelectric mechanism can typically convert a maximum of ∼4 % of the FUV energy into gas heating. This includes the assumption of a ∼ 10% probability that an absorbed photon results in an ejected electron (the photoelectric yield). Wright et al. (1991) determined, based on *COBE* data, that the 158 μm C$^+$ line alone emits 0.3% of the bolometric far infrared (≈ UV) luminosity of our Galaxy.

The UV flux incident on a neutral cloud, $G_0$, is usually expressed in units of the Habing (1968) flux of $1.6 \times 10^{-3}$ ergs cm$^{-2}$ s$^{-1}$. For unusually highly excited regions, such as Orion, the UV flux incident on the neutral PDR region is $G_0 \sim 10^{4-5}$ and the gas density is $n \sim 10^5$ cm$^{-3}$. The ionization potential of carbon is 11.3 eV, so carbon can be singly ionized where hydrogen is neutral. The ionization potential of oxygen is 13.6 eV, like that of hydrogen, so oxygen is neutral where hydrogen is neutral. In general, for high UV fields ($G_0 > 10^{2.5}$) and/or high densities ($n > 10^{3.5}$ cm$^{-3}$), the O$^0$ fine structure lines at 63 μm and 145 μm contribute significantly to the cooling of the neutral gas. For lower densities and UV field strengths, the 157.741 μm (Cooksy, Blake, & Saykally 1986) transition of C$^+$ ($^2P_{3/2} - ^2P_{1/2}$) is the major cooling line. This is because in O$^0$ the $^3P_1$ excited state is 228 K above the $^3P_2$ ground state, and the $^3P_0$ state is 326 K above the ground state. In C$^+$ the $^2P_{3/2}$ excited state is only 91 K above the ground state. Dust grains become positively charged, especially near the surfaces of clouds, resulting in a decrease of the photoelectric heating efficiency. An increased electron density results in a higher dust-electron recombination rate, allowing the UV energy to be converted to gas heating more efficiently. Thus, the influence of an increase in density is similar to an increase in the gas excitation.

The visual extinction into a neutral cloud, $A_V$, is related to the hydrogen column density, $N_H$, by $N_H/A_V = 1.8 \times 10^{21}$ cm$^{-2}$ mag$^{-1}$ (cf. Burstein, & Heiles 1978, Bohlin, Savage, & Drake 1978), although this is somewhat dependent on the local gas-to-dust ratio. Hydrogen in the PDR is predominantly in the form of H I for $A_V < 2$ and H$_2$ for $A_V > 3$. C$^+$ dominates the cooling of the outer regions of the cloud ($A_V < 4$) for low-excitation PDRs, as described by Hollenbach, Takahashi, & Tielens (1991). C$^0$ should dominate the cooling for $4 < A_V < 6$ and CO rotational transitions dominate the cooling for $A_V > 6$. The $A_V$ values, above, vary according to the detailed physical properties of the particular PDR (i.e. $n$, $G_0$, and geometrical effects). Significant UV shielding (i.e. $A_V \sim 10 - 20$) is required before OH, H$_2$O, and O$_2$ can be sustained. For $A_V > 6$ heating from cosmic rays and gas-grain collisions become significant relative to the UV heating.

The photodissociation model has been successfully applied to Orion, which is a face-on PDR (Tielens & Hollenbach 1985b), and to M17SW, which is an edge-on PDR (Meixner 1992). Wolfire, Tielens, & Hollenbach (1990) applied the model to the nuclear region of our Galaxy and to M82 to derive UV field strengths and gas densities. Schilke et al. (1993) also examined M82, but they find that the neutral atomic carbon emission is anomalously enhanced above the predicted levels from the standard PDR model. This may be due to an unusually high cosmic ray rate in M82. Low density PDR models, which are important for interpreting the average PDR conditions in our Galaxy, were computed by Hollenbach, Takahashi, & Tielens (1991). Wolfire, Hollenbach, & Tielens (1992) conclude that CO 1-0 emission from giant molecular clouds is primarily the result of the PDR physics. Crawford et al. (1985) and others have found that PDR models account for the excellent correlation both between 158 $\mu$m C$^+$ emission and CO 1-0 emission, and between C$^+$ and far IR continuum emission.

## 2. Observations & Previous Results

Results are presented from the *COBE* Far InfraRed Absolute Spectrophotometer (FIRAS) instrument. The instrument is a scanning Michelson interferometer that covers the wavelength range from 100 $\mu$m to 1 cm. While the primary purpose of the instrument is to make precise measurements of the cosmic microwave background radiation, the instrument design offers many advantages for the study of large scale diffuse emission from our Galaxy. For example, the FIRAS has a 7° beam, has the ability to make absolute measurements of intensity with no beam switching, and has observed nearly the entire sky. For these reasons, the FIRAS is well-suited for measurements of diffuse emission. The FIRAS spectral resolution is low, as discussed below in §3, and no spectral lines are resolved. Despite this, the general galactic rotation can be discerned using the detected FIRAS spectral lines.

The first cosmological results from the *COBE* FIRAS instrument were reported by Mather et al. (1990), and recent overall summaries of *COBE* results are found in Boggess et al. (1992) and Bennett et al. (1992a,b). More recent cosmological results are reported by Mather et al. (1994), Fixsen et al. (1994a), and Wright et al. (1994).

The first results of the *COBE* FIRAS unbiased far infrared survey of the Milky Way were published by Wright et al. (1991). Table 1 in that paper summarizes spectral line detections from *COBE* FIRAS: 158 $\mu$m C$^+$, 370 $\mu$m and 609 $\mu$m C$^0$, J=2-1 through J=5-4 CO, and 122 $\mu$m and 205 $\mu$m N$^+$. Since that time, improved data processing algorithms and techniques have allowed for a more detailed examination of the data. In addition, since the time of the Wright et al. (1991) analysis, a laboratory detection has been reported for the 205 $\mu$m ground state transition in N$^+$. The frequency of the N$^+$ transitions are measured to be 1461.1319(6) GHz and 2459.3801(4) GHz (Brown & Evenson 1993).

The $C^+$, $C^0$, and CO emission lines detected by *COBE* can immediately be recognized as the expected cooling lines from PDRs with low densities and small UV fields (Hollenbach, Takahashi, & Tielens 1991). The result that the galactic-averaged PDR properties are described by lower excitation conditions than objects such as Orion is reinforced by the weakness of the 145 $\mu$m $O^0$ line.

The two transitions of $N^+$ are recognized as arising from the ionized medium. As pointed out by Wright et al., this emission can not be coming only from classical, high-density, H II regions. The ratio of the intensities of the $^3P_2$ to $^3P_1$ 122 $\mu$m $N^+$ line to the $^3P_1$ to $^3P_0$ 205 $\mu$m ground state transition is a probe of the density of the ionized gas. Rubin (1985) provides a grid of classical H II region models that result in $I(N^+\ 122\ \mu m)/I(N^+\ 205\ \mu m)$ ranging from 3 (for $n_e \approx 10^2$ cm$^{-3}$) to 10 (for $n_e \approx 10^5$ cm$^{-3}$). In the low-density limit, $n_e \ll 100$ cm$^{-3}$, the expected intensity ratio is 0.7. Wright et al. reported that the measured *COBE* ratio is 1.0 to 1.6, depending on the adopted values for the instrumental resolution at the two wavelengths. Thus the ionized gas measured by *COBE* FIRAS is likely to arise mostly from large, diffuse, regions. Reynolds (1990, and references therein) has extensively studied the warm ionized medium (WIM), concluding that a large fraction of the Galaxy contains $\sim$ 8000 K fully ionized gas with a density of $n_e \approx 0.2$ cm$^{-3}$. However, the galactic disk is opaque to the H$\alpha$ and other optical lines used to study this medium so only a local (typically < 2 kpc) view exists. The galactic disk is transparent to microwave continuum free-free emission, but it is difficult to separate the dust and synchrotron components from the overall microwave emission (Bennett et al. 1992c). Radio recombination line observations have not been sensitive enough to observe the WIM.

## 3. Line Fitting and Calibration

The FIRAS instrument and its calibration were described in detail by Fixsen et al. (1994b). In this paper we use data from all four detectors (Left Low, Left High, Right Low, and Right High) and all of the interferometer mirror scan modes used in flight (Short Slow, Short Fast, and Long Fast). Data from all of the channels, scan modes, and frequency bands were combined using a weighted average to form a single data set with 167 spectral points from 4.4 mm to 104 $\mu$m (2.3 to 96.3 cm$^{-1}$). Right High Short Slow (RHSS) mode data dominate the high frequency spectrum (20-96 cm$^{-1}$) and Left Low Short Slow (LLSS) mode data dominate the low frequency spectrum (2-20 cm$^{-1}$).

The ideal instrument line response function for an interferometer is

$$I(x) = \Omega_{beam}^{-1} \int d\Omega \int_0^\infty d\nu \cos(2\pi\nu x \cos\theta + \phi), \tag{1}$$

where $x$ is the path length difference in cm between the two arms of the interferometer, $\nu$ is the line frequency in waves per centimeter, $\theta$ is the angle of each ray from the beam axis, $d\Omega$ is a differential of the solid angle of the beam, $\phi$ is the phase shift of the interferometer at the frequency $\nu$, and $\Omega_{beam}$ normalizes the solid angle integral over the beam. The instrument produces interferograms that we Fourier transform to recover the approximate sky spectrum,

$$S(\nu') = \int_{-\infty}^\infty dx\, e^{2\pi i \nu' x} A(x) I(x) \tag{2}$$

or

$$S(\nu') = \Omega_{beam}^{-1} \int_{-\infty}^\infty dx\, e^{2\pi i \nu' x} A(x) \int d\Omega \int_0^\infty d\nu \cos(2\pi\nu x \cos\theta + \phi). \tag{3}$$

The factor $A(x)$, the apodization function, is a Fourier window function that accounts for the fact that data are not obtained with equal weight over the full range of the integral.

With $A(x) = 1$, the integral over solid angle (with cylindrical symmetry) produces a rectangular spectral line profile of width $\Delta\nu/\nu = 1 - \cos(\theta_{max})$, where $\theta_{max} = 7°$ is the cut-off angle of the FIRAS interferometer. The effect of this cut-off angle can be represented in terms of an additional Fourier window, or self-apodization function, $A_{self}(x)$, while taking $\cos(\theta_{max}) = 1$. Ideally $A_{self}(x)$ would be the Fourier transform of a rectangular spectral line profile, i.e. a sinc function, but in practice there is some taper in the instrument profile that makes a Gaussian a better approximation than a sinc function. For the FIRAS instrument $\Delta\nu/\nu = 0.007$ FWHM by design, but the measured $A_{self}$ is approximately Gaussian with a FWHM of $\Delta\nu/\nu = 0.0035$ at the $C^+$ line. A possible explanation for the discrepancy is that the expected large optical aberrations reduce the optical efficiency near the edge of the beam.

For the FIRAS instrument, $A(x) = 0$ where data are not taken, so the instrument line profile width is further broadened. Combining this effect with the self-apodization function, we take the total $A_{tot}(x) = A(x) A_{self}$. We take $A(x)$ to be a particular asymmetric smooth window function to enhance the signal-to-noise ratio for wide spectral features and minimize ringing from spectral lines. The Fourier transform of $A(x)$ is the dominant factor in determining the instrument line profile. The piecewise expression for the apodization function $A(x_i)$ for the 512 sampled data points is

$$A_i = f_i \left[1 - \left(\frac{i-c}{j-c}\right)^4\right]^2 \quad \text{for } i \in [1, 512] \tag{4}$$

where $f_1 = f_2 = 0$. For Left and Right Low Long Fast (LLLF and RLLF) data $j = 513$, $c = 88$, and

$$f_i = \begin{cases} 0 & i \in [1, 2] \\ \{1 - \cos\left(\frac{\pi(i-2)}{30}\right)\}/2 & i \in [3, 32] \\ 1 & i \in [33, 2c - 32] \\ \{3 - \cos\left(\frac{\pi(i+32-2c)}{30}\right)\}/2 & i \in [2c - 31, 2c - 2] \\ 2 & i \in [2c - 1, 512]. \end{cases} \quad (5)$$

For all other scan mode data $j = 1$ and

$$f_i = \begin{cases} 0 & i \in [1, 2] \\ 2 & i \in [3, 2c - 513] \\ \{3 - \cos\left(\frac{\pi(2c-482-i)}{30}\right)\}/2 & i \in [2c - 512, 2c - 483] \\ 1 & i \in [2c - 482, 482] \\ \{1 - \cos\left(\frac{\pi(513-i)}{30}\right)\}]/2 & i \in [483, 512], \end{cases} \quad (6)$$

(as seen in Figure 7 of Fixsen et al. 1994b), where for the Right and Left High Long Fast (RHLF and LHLF) modes $c = 353$, for all Short Slow scan modes $c = 355$, and for Right and Left High Short Fast (RHSF and LHSF) modes $c = 356$.

Since $A(x)$ is asymmetric, the resulting line profile is complex. The real (complex) part of the transform of $A(x)$ has a width of 0.65 cm$^{-1}$ (0.52 cm$^{-1}$) for Short mirror scan mode data and 0.16 (0.13 cm$^{-1}$ (0.13 cm$^{-1}$) FWHM for Long mode data (usable only for frequencies below 20 cm$^{-1}$). Thus the instrument spectral profile model is generated from the ideal cosine wave interferogram with the two window functions applied, $A_{tot}(x) = A_{self}(x)A(x)$. The resulting spectrum is calibrated, which involves dividing it by a frequency-dependent gain function. The final line profile is asymmetric and complex.

Continuum and line spectra were fitted to the data as follows. A 2.726 K Planck spectrum was subtracted from the spectrum at each sky pixel. Complex parameters of the fit (and their corresponding degrees of freedom at each pixel) were: the amplitudes at each assumed line frequency (32), the amplitude of a d(Planck)/dT term to remove the cosmic microwave dipole (2), and coefficients of Legendre polynomials for the continuum emission of the dust (40). The fit removes 74 degrees of freedom, leaving 260 degrees of freedom at each of 5856 pixels. The data were weighted by an inverse variance that is dominated by detector noise, but includes other error terms (see Fixsen et al. 1994b). The use of this diagonal matrix weighting is an approximation to the full covariance matrix, which is not perfectly diagonal because of the effects of the apodization and the calibration process. Although we have included the imaginary parts of the data in our analysis we will discuss only the physically meaningful real results, while the imaginary results provide useful information on the errors.

## 4. Analysis Results

### 4.1. The Galactic Plane

In the Wright et al. (1991) analysis the dust continuum emission map of the sky was used as a spatial filter for averaging the emission from the spectral lines, effectively assuming that the spectral line intensities follow the continuum spatial distribution. That assumption is not made here. Rather, for each pixel we perform a simultaneous least squares fit of a continuum spectral "baseline" plus a series of spectral line profiles centered on the wavelengths of known lines. The result is a set of spectral line intensities for each sky position. We find that while the $C^+$ emission (Figure 1, top) follows the continuum closely, and the $N^+$ emission (Figure 1, bottom) follows it approximately, the $C^0$ and CO emission are distinct from the continuum template used by Wright et al., in that the $C^0$ and CO flux distributions are strongly concentrated towards the galactic center.

In Figure 2a the longitude profiles of the $C^+$ and 158 $\mu$m far infrared continuum emission are shown for the entire galactic plane. The line and continuum emission are seen to track one another very well, as would be expected for PDR gas. The emission appears to peak towards the molecular ring, the galactic center direction, and in the spiral arms. Only towards the galactic center does the $C^+$ emission appear to be relatively weak compared with the far infrared continuum. In contrast, the 370 and 609 $\mu$m $C^0$ emission and the CO emission, shown in Figure 2 (i-p), peak strongly towards the galactic center direction. The J=2-1, 3-2, 4-3, and 5-4 lines are detected, while upper limits are placed on the CO J=1-0 and 6-5 transitions.

Figure 2c and 2d show the profiles of the two $N^+$ transitions. Although the 122 $\mu$m transition is noisier due to the reduced instrument sensitivity at that wavelength, the transitions clearly track one another, as expected. If we calculate a constant 122 to 205 $\mu$m intensity ratio we arrive at $0.9 \pm 0.1$, consistent with the lower value given by Wright et al. (1991). However, the derivation of the intensity ratio of 0.9 is dominated by the stronger regions, and is not a particularly good fit overall. If, in addition to fitting for a constant ratio term, we include a term proportional to the 205 $\mu$m $N^+$ line intensity of the form $I(N^+\ 122\ \mu m)/I(N^+\ 205\ \mu m) = 0.7 + I(N^+\ 205\ \mu m)/I_0$, then we find that the preferred solution is $I_0 = 7 \times 10^{-5}$ erg s$^{-1}$ cm$^{-2}$ sr$^{-1}$. That is, in the regions where the 205 $\mu$m $N^+$ emission is strongest the intensity ratio is about unity, but in weaker regions the intensity ratio is indistinguishable from the 0.7 ratio expected for regions of low electron density.

The spatial distributions of $C^+$ and $N^+$ are unmistakably related, but the $N^+$ intensity increases proportionally more than the continuum intensity while the $C^+$ intensity follows the continuum in the plane of the Galaxy. Figure 3 shows that an excellent correlation exists between the fluxes of $C^+$ and $N^+$, in the sense that $I(N^+) \propto I(C^+)^{1.5}$, as pointed out by Petuchowski & Bennett (1993a,b) and Bennett & Hinshaw (1993). They suggest that the 1.5 power law may be the result of a volume to surface area ratio sampling of the interstellar gas. Figure 4 shows that $I(C^+) \propto I(\text{FIR})^{0.95}$ while $I(N^+) \propto I(\text{FIR})^{1.53}$, where the far IR intensity, $I(\text{FIR})$, is defined as the continuum intensity at each respective wavelength within the same effective bandwidth as the line.

The 1.5 index power-law may be suggestive of a volume to surface area ratio geometric effect. In such a scenario the volumes of very large internally ionized regions give rise to $N^+$ emission. These regions are partially surrounded by neutral gas with the interface surface traced by $C^+$ emission. If the emission from each line of sight is dominated by one such region that is characterized by an effective radius $R_{eff}$, then $I(N^+) \propto R_{eff}^3$ and $I(C^+) \propto R_{eff}^2$, so from one line of sight to another the ratio will be characterized by $I(N^+)/I(C^+) \propto R_{eff}^{1.5}$. For the 1.5 power law to hold each line of sight must be better characterized by an effective radius than by the number of regions along each line of sight. If several regions contribute substantially to the emission along each line of sight, then $I(N^+)$ would be linearly proportional to $I(C^+)$. In this picture, the emission, which peaks in the direction of the molecular ring, is dominated by approximately beam-sized bubbles of ionized gas surrounded by neutral gas. For the *COBE* 7° beam, this corresponds to bubbles 1 kpc in diameter. Comparable size bubbles are known to exist in external galaxies (e.g. Bruhweiler, Fitzurka & Gull 1991; Kamphius, Sancisi & van der Hulst 1991). Note that the $N^+$ emission need not arise from the volume of the bubble, but could be from swept up gas so long as the amount of gas is characterized by the bubble volume.

The above scenario requires large-scale structures. Shibai et al. (1991) report results from a 3.4 arc-minute beam size survey of $C^+$ emission concluding that, "most [C II] emission of the Galactic disk is radiated not from discrete sources but from more extended and diffuse regions." Nakagawa et al. (1993) have shown $C^+$ maps made from observations with a 12 arc-minute beam that also shows structures that are much larger than the beam size. Doi et al. (1993) conclude that discrete sources are significant when looking through the length of the $l$=90° spiral arm.

Massive and luminous O and B stars tend to form in clusters and then become Type II supernovae that are correlated in space and time. The correlated supernovae create bubbles in the plane of the Galaxy. If a bubble is large enough it can break through the thickness of the disk. If the break-through is particularly energetic it will blow material into the halo of the Galaxy. These blowouts lead to Galactic fountains. In these models significant amounts of material from the galactic plane are thrown out of the plane and at least some of this material rains back down onto the plane. These processes have been described by Bregman (1980), Houck & Bregman (1990), Mac Low & McCray (1988), Mac Low, McCray, & Norman (1989), Heiles (1990), Norman & Ikeuchi (1989), and Shapiro & Field (1976). Heiles (1990) estimates an area filling factor of breakthrough bubbles of 23% in our Galaxy. McKee & Williams (1993a,b) derive the somewhat lower filling factor of superbubbles of 10%. We feel that the cycle involving giant molecular clouds, massive clustered O and B stars, and clustered supernovae must be greatly enhanced in the molecular ring so that the filling factors of bubbles, quoted above, are probably underestimates for the ring.

Petuchowski & Bennett (1993a, 1993b) examined the possible make-up of the *COBE* FIRAS line emission in terms of three morphologies: classical ionization bounded H II regions, boundary ionized clouds, and an extended low density warm ionized medium. They conclude that the classical ionization bounded H II regions contribute very little to the total line emission intensities, largely because of their small volume filling factors. The other two morphologies can occupy large filling factors and contribute significantly to the total signal intensity. Since bubbles, superbubbles, and the neutral material around their periphery, are examples of large filling factor regions, they are capable of dominating observed signals along lines of sight that also contain several, much smaller, classical H II regions.

Volume-to-surface geometry is only one possible explanation for the 1.5 power scaling law between $C^+$ and $N^+$. It may be that the $C^+$ scales proportionally with the far infrared continuum for the average PDR conditions in the Galaxy, but that $I(N^+) \propto I(\text{FIR})^{\sim 1.5} \propto I(\text{UV})^{\sim 1.5}$ for some other physical reason, for example, having to do with the variations of color of the UV radiation field in the Galaxy. It is worth noting that a similar 1.5 power law relation has been reported for different transitions in an external galaxy (Petuchowski & Bennett 1993a,b).

H I (c.f. Dickey & Lockman 1990) and $C^+$ *appear* to have different spatial distributions, but this does not contradict the picture that $C^+$ cools H I gas. Except where the emission is self-absorbed, the H I flux is directly proportional to the H I column density. On the other hand, the 158 $\mu$m $C^+$ flux does not, in general, scale directly with the gas density. Since $C^+$ emission is the result of a collisional process, one might naively expect $C^+$ emission to scale as the square of the gas density. In the low excitation PDR models of Hollenbach, Takahashi, & Tielens (1991), the 158 $\mu$m $C^+$ emergent intensity scales somewhat less than the square of the density for low densities and low UV field strengths. Nevertheless, since H I and $C^+$ intensities scale differently with density the fact that they have *apparently* different spatial distributions is not surprising. In addition, $C^+$ cools a fraction of the $H_2$ gas (to a depth of $A_V \approx 4$) as well as the H I gas, and warm H I gas has additional coolants available, further complicating comparisons of $C^+$ and H I distributions.

As discussed above, the 145.5 $\mu$m line of $O^0$ is a potentially important PDR cooling line. Figure 2e shows the galactic plane distribution of the 145.5 $\mu$m line of neutral atomic oxygen, which Wright et al. (1991) listed as a nearly $2\sigma$ feature. The mean ratio to the $C^+$ intensity is $I(C^+)/I(O^0) = 71 \pm 36$, assuming a single intensity scaling factor. We caution that the 145.5 $\mu$m $O^0$ line is close in wavelength to the very much stronger 158 $\mu$m $C^+$ line so that an error in our assumed instrument line response to $C^+$ at 158 $\mu$m of only 2% at 145.5 $\mu$m could produce the $O^0$ feature.

Figure 2 also shows the galactic plane solutions for other selected transitions in the FIRAS wavelength range.

## 4.2. The Galactic Center ($< 500$ pc) Region

The physical properties of the Galactic Center (GC) region gas are unlike those elsewhere in the Galaxy. This is likely a result of the steep gravitational potential gradient within 500 pc of the GC where there are $10^8$ $M_\odot$ of molecular gas (Genzel 1992; Güsten 1989). For molecular clouds near the GC to be stable against tidal disruption their mean $H_2$ density must be about $10^4$ cm$^{-3}$ or more (Güsten & Downes 1980), in contrast to the typical disk molecular cloud density of $\approx 10^2$ cm$^{-3}$ (Scoville & Sanders 1987). Density-sensitive measurements of CS 1-0/2-1 show extensive high density regions within the central 500 pc (Tsuboi et al. 1989; Stark et al. 1989). The volume filling factor of molecular gas in the GC is two orders of magnitude higher than in the plane, resulting in mean gas densities of 100 cm$^{-3}$ in the GC versus 1 cm$^{-3}$ in the plane (Sanders, Solomon, & Scoville 1984; Scoville & Sanders 1987). While individual clouds near the GC may be close to virialization, the larger cloud complexes are not (Stark & Bania 1986). The gas temperatures near the GC (40-150 K) are much higher than the dust temperatures ($< 30$ K), as well as being higher than the temperatures of a typical disk molecular clouds (10 K) (Bally, Stark & Wilson 1987, Bally, Stark & Wilson 1988, Morris et al. 1983, Liszt & Burton 1978). Collisional heating of the gas by radiatively heated dust thus appears not to be the dominant heating mechanism near the GC. Blitz et al. (1985) report a decrease in the inferred cosmic ray rate toward the GC, based on COS-B $\gamma$-ray data, so a high rate of cosmic ray heating also appears to be ruled out. Heating by young stars appears to be ruled out as well because of the low dust temperatures and other indications of a low level of active star formation (Morris 1989; Cox & Laureijs 1989). The high gas densities themselves appear to rule out magnetic heating effects. The unusually large gas velocity-induced linewidths observed near the GC suggest the possibility of heating by turbulent energy dissipation (Wilson et al. 1982; Güsten et al. 1985). Since the turbulent energy dissipation time constant is short ($\sim 10^6$ years), new energy must be provided almost continuously, perhaps from the strong differential rotation in the central bulge (Fleck 1980).

The AT&T Bell Labs CO J=1-0 survey (Bally, Stark & Wilson 1987) shows a region that extends over galactic longitudes $-1° < \ell < +2°$ and about $0.4°$ of galactic latitude (i.e. about 400 pc × 50 pc in size). This region appears as a parallelogram in a $\ell v$ diagram, indicating that much of the neutral material is in nearly rigid-disk rotation about the galactic center. Several H II regions detected by the H109$\alpha$ radio recombination line indicate star formation activity has occurred there in the last $10^8$ years, and the positions of the H II regions suggest that star formation occurred in a 100 pc ring about the galactic center.

Caution must be used in using CO to trace $H_2$ gas. For spherically symmetric neutral clouds in virial equilibrium with a radial velocity field ($v \propto r$) the CO luminosity is given by $L_{CO} = (3\pi G/4\rho)^{0.5} T_{CO} M$ (c.f. Scoville & Sanders 1987), where $G$ is the gravitational constant, $\rho$ is the cloud mass density, $T_{CO}$ is the peak brightness temperature in the CO transition, $M$ is the cloud mass, and $L_{CO}$ is in the usual units of K km s$^{-1}$ pc$^2$. The ratio of $L_{CO}$ to $M$ is constant only as long as the geometry, dynamics, and composition of the clouds are constant, and $T_{CO}/\rho^{1/2}$ is constant. Since temperature and density often rise together, this leads to a generally robust estimate that $X_{CO} \equiv N_{H_2}/I_{CO} \simeq 3.6 \times 10^{20}$ cm$^{-2}$ (K km s$^{-1}$)$^{-1}$. However, many assumptions go into the CO to $H_2$ conversion above. At one extreme, a molecular cloud can have all of its carbon in the form of $C^0$ and $C^+$ and no CO, but with $H_2$ at its core. In this case the CO intensity (zero!) underestimates the $H_2$ mass. At the other extreme a large dense molecular cloud with only a small surface thickness of $C^+$ can have almost all of its carbon in the form of CO. The fact that different volume-to-surface area ratios exist between molecular clouds in the disk and near the GC implies that one might expect a different CO to $H_2$ conversion factor.

The far infrared fine structure lines, as seen by the *COBE* FIRAS, towards the GC are unlike any other region of the sky, and deserve special comment. 3.5° corresponds to 500 pc at the GC, assuming a distance of 8.5 kpc. While there is a local maximum of emission of $C^+$ towards the GC, the $C^+$ emission has a global emission maximum towards the molecular ring. On the other hand, the CO and $C^0$ transitions peak strongly in the GC direction. Thus, the cooling of the GC neutral gas is being done proportionally more by the CO and $C^0$ forms of carbon than by $C^+$. This would occur if the neutral gas towards the GC has a larger volume-to-surface area, or in other words, is better shielded, than neutral gas in the disk. Therefore, based on the observations of the *COBE* FIRAS carbon cooling lines, we suggest that there is an unusually large amount of well-shielded dense neutral gas in the GC direction compared with other lines of sight. Since the GC neutral material is of higher density, relatively more cooling is expected from $C^0$ and CO than by $C^+$ because of the higher relative filling factor of high-$A_V$ neutral gas.

Alternately, differences in the FIRAS observations of the cooling lines towards the GC could be indicative of other effects such as a saturation of the 158 $\mu$m $C^+$ line towards the GC or self-absorption of $C^+$ emission towards the galactic center, but the 1.5 power law scaling with the $N^+$, discussed above, still holds in this region.

The ground state $0_{00} - 1_{11}$ 269 $\mu$m transition of para-$H_2O$ has never been detected astronomically. This is largely due to the enormous opacity of the Earth's atmosphere at this wavelength. The longitude profile of this water transition, shown in Figure 2h, displays an absorption-like dip towards the galactic center, although the signal-to-noise ratio is inadequate for us to be able to claim a detection of this line. The intensity dip at 269 $\mu$m is $\sim 2.5\sigma$ in frequency space. With about 5000 sky samples one expects 38 such dips on the sky at this wavelength and, in fact, there are 38. However, only towards the galactic center are three approximately independent dips so spatially close to one another. The error bars on the line intensity are larger near the galactic center because of uncertainty in the ability to calibrate and remove the larger continuum emission there. Thus we consider the water intensity dip suggestive, but not conclusive, and pose the question of whether such an absorption would be astrophysically reasonable.

Water has a dipole moment of $\mu_D = 1.9$ Debye, corresponding to the 269 $\mu$m para-water transition rate $A = 1.86 \times 10^{-2}$ s$^{-1}$. If we assume the abundance of water is $[H_2O]/[H_2] = 10^{-6}$, that 10% of all water resides in the gaseous ground state, and that the density of an absorbing cloud is typically $10^2$ cm$^{-3}$, then the spectral line will be very optically thick. Hence, we expect optically thick water absorption in this transition to be common. For a linewidth corresponding to a velocity dispersion of 30 km s$^{-1}$ (a typical width for lines near the GC), the FIRAS observation would suffer a frequency dilution factor of 150. Taking the mean value of the negative intensity dip as a real absorption gives $\Delta I = 6 \times 10^{-7}$ erg s$^{-1}$ cm$^{-2}$ sr$^{-1}$ with a continuum intensity at that location and wavelength of $I = 7 \times 10^{-5}$ erg s$^{-1}$ cm$^{-2}$ sr$^{-1}$, so $\Delta I/I = 9 \times 10^{-3}$. Correcting for the frequency filling factor gives $150 \Delta I/I \approx 1$, so the measurement is consistent with optically thick water absorption. The corresponding ground state transition of ortho-water is at 538 $\mu$m, also observable by FIRAS. This transition will also be optically thick, so any absorption feature in 269 $\mu$m para-water should appear at 538 $\mu$m at approximately a fifth of the 269 $\mu$m intensity. There is no evidence for any absorption dip at 538 $\mu$m (see Figure 2g), but the expected level of $\Delta I = 10^{-7}$ erg s$^{-1}$ cm$^{-2}$ sr$^{-1}$ is below the level of the noise. Thus, while we can not claim a detection of this water line, the intensity dip observed at 269 $\mu$m is consistent in wavelength, spatial position, and amplitude with what is expected astrophysically.

The *Submillimeter Wave Astronomy Satellite (SWAS)* will observe the ortho-water ground state transition at 538 $\mu$m, in addition to its analog in $H_2^{18}O$ at 547 $\mu$m, 609 $\mu$m $C^0$, 544 $\mu$m $^{13}CO$ J=5-4, and 615 $\mu$m $O_2$. These lines fall in, or very near, the break between the high and low frequency FIRAS bands where the instrument response is greatly reduced. The 705.8 $\mu$m transition of $O_2$ is not detected by *COBE* FIRAS, and we place a limit on its intensity towards the galactic center of $2 \times 10^{-7}$ erg s$^{-1}$ cm$^{-2}$ sr$^{-1}$. We place a limit of $4 \times 10^{-7}$ erg s$^{-1}$ cm$^{-2}$ sr$^{-1}$ on the intensity of the 538.3 $\mu$m transition of $H_2O$ towards the galactic center.

### 4.3. High Galactic Latitude Results

Only the 158 $\mu$m line of $C^+$ and the 205 $\mu$m line of $N^+$ are sufficiently strong to allow useful estimates of their intensities at high galactic latitudes.

We compute the ratio of the 158 $\mu$m $C^+$ intensity to the H I column density for the entire sky at 7° resolution. Figures 5 and 6 show the relationship between the *COBE* 158 $\mu$m $C^+$ emission and H I column density (Dickey & Lockman 1990). A fit to the H I data gives a plane parallel approximation of $N(H\ I) = 3.3 \times 10^{20}$ cm$^{-2}$ csc $|b|$ (Lockman, Hobbs, & Shull 1986). The difficulty in fitting for the $C^+$ cooling rate from the data in the low intensity portion of Figure 6 (*top*) is that not all of the $C^+$ emission comes from regions coextensive with atomic H I emission. $C^+$ emission also arises from molecular $H_2$ and ionized hydrogen regions. For regions where significant $H_2$ gas exists data points in Figure 6 are effectively shifted to the left, since the H I column density underestimates the total neutral hydrogen along the line of sight. For ionized hydrogen regions with significant $C^+$ emission, the data points in Figure 6 effectively take on a positive vertical displacement. Thus Figure 6 should contain an underlying linear relation with additional "contaminated" data above the linear relation. We can limit the contamination by selecting only data greater than 30° from the galactic center with $N(HI) < 6 \times 10^{21}$ cm$^{-2}$. This eliminates most of the data where CO plays a major role in the cooling of the gas (i.e. where the hydrogen gas is in molecular $H_2$ form), and much of the gas where hydrogen is ionized in high column density lines of sight. A simple linear regression to the remaining data, shown in the lower panel of Figure 6, gives a cooling rate of $4\pi I(C^+)/N(H\ I) = 2.8 \times 10^{-26}$ erg s$^{-1}$ per H-atom. However, these data remain clearly contaminated. Much of this contamination arises from the Orion and Gum Nebulae. We iteratively perform a median robust estimation of linear slope and intercept coefficients and then remove the data that most deviates from the line. This results in a cooling rate of $2.65 \times 10^{-26}$ erg s$^{-1}$ per H-atom. The uncertainty in removing the "contamination" dominates, by far, the *COBE* FIRAS data uncertainties. We believe a "best" value for the diffuse C$^+$ cooling rate is $4\pi I(\mathrm{C}^+)/N(H\,I) = (2.65 \pm 0.15) \times 10^{-26}$ erg s$^{-1}$ per H-atom. The contribution of C$^+$ emission from the ubiquitous Warm Ionized Medium can be estimated from the $I(\mathrm{C}^+)$-intercept of the plot in Figure 6. We place a limit of

$$\lim_{N(HI)\to 0} I(\mathrm{C}^+) < 6 \times 10^{-7}\ \mathrm{erg\ cm}^{-2}\ \mathrm{s}^{-1}\ \mathrm{sr}^{-1}(95\%\ CL) \qquad (7)$$

Bock et al. (1993) conclude that at a high galactic latitude the C$^+$ cooling rate is $(2.6 \pm 0.6) \times 10^{-26}$ erg s$^{-1}$ per H-atom in the small region of sky that they sampled and claim a detection of a nonzero intercept from ionized gas of

$$\lim_{N(HI)\to 0} I(\mathrm{C}^+) = (9 \pm 3.5) \times 10^{-8} \mathrm{erg\ cm}^{-2}\ \mathrm{s}^{-1}\ \mathrm{sr}^{-1}. \qquad (8)$$

The agreement of the cooling rates measured by Bock et al. with high resolution over a small sky region and our results with low resolution over nearly the full sky is especially reassuring. A UV-absorption derived C$^+$ cooling rate of $2.1 \times 10^{-26}$ erg s$^{-1}$ per H-atom was reported by Savage et al. (1993), less than the earlier result of Pottasch, Wesselius & van Duinen (1979) of $1.3 \times 10^{-25}$ ergs s$^{-1}$ per H-atom. Gry, Lequeux, & Boulanger (1992) reported a C$^+$ cooling rate of $3.5^{+5.4}_{-2.1} \times 10^{-26}$ ergs s$^{-1}$ per H-atom, but their suggestion that the *COBE* C$^+$ emission may come mainly from H II regions appears to be incorrect.

The spin-orbit interaction removes the ground state degeneracy in singly ionized carbon causing the 158 $\mu$m magnetic dipole transition. Thus we consider C$^+$ as a two-level system and, denoting the $^2\mathrm{P}_{1/2}$ ground state as level "1" and the $^2\mathrm{P}_{3/2}$ first excited state as "2", the critical density for level 2 is defined as $n_{cr} \equiv A_{21}/q_{21}$, where $q_{21} \equiv 8.629 \times 10^{-6}\Omega/g_2 T^{1/2}$ cm$^3$ s$^{-1}$ is the collisional de-excitation rate and $\Omega$ is a constant. For C$^+$-e$^-$ collisions we adopt $\Omega = 0.444$ (Hayes & Nussbaumer 1984), replacing the earlier laboratory result (1.34 of Tambe 1977). For C$^+$-H collisions we adopt $\Omega = 0.00292$ (Launay & Roueff 1977). The collisional excitation and de-excitation rates are related by $q_{12} = (g_2/g_1)q_{21}e^{-h\nu/kT}$. The cooling function, $L_i$ (erg cm$^3$ s$^{-1}$), resulting from collisions with partner "i" is given by $L_i = (q_{12})_i h\nu = (g_2/g_1)(q_{21})_i h\nu e^{-h\nu/kT}$.

If the 158 $\mu$m line originates in neutral gas then its intensity is given by (c.f. Crawford et al. 1985)

$$I(\mathrm{C}^+\ 158\mu\mathrm{m}) = \frac{1}{4\pi} h\nu A \frac{[\mathrm{C}^+]}{[\mathrm{H}(\mathrm{C}^+)]} N_H(\mathrm{C}^+)\Phi_b \frac{g_2}{g_1} e^{-h\nu/kT} \left\{1 + \frac{g_2}{g_1}e^{-h\nu/kT} + \left[\sum_i \left(\frac{n}{n_{cr}}\right)_i^{-1}\right]^{-1}\right\}^{-1} \qquad (9)$$

where the summation runs over the various significant collision partners that excite $C^+$ (e.g. H I, H$_2$, and electrons), $h\nu = 1.26 \times 10^{-14}$ ergs, the transition rate, $A = 2.36 \times 10^{-6}$ s$^{-1}$, the statistical weights of the states are $g = 2J + 1$ ($g_2 = 4$, $g_1 = 2$), and we assume a beam filling factor of $\Phi_b = 1$. $N_H(C^+)$ is the column density of neutral hydrogen that is coextensive with the $C^+$ emission. The critical density, $n_{cr}$ depends on the the $C^+$ excitation partner. For collisions with H I, $n_{cr}(H\ I) \sim 3000$ cm$^{-3}$, and for collisions with H$_2$, $n_{cr}(H_2) \sim 4000$ cm$^{-3}$. For electron collisions $n_{cr}(e^-) \sim 10$ cm$^{-3}$. If we assume that all of the emission results from coextensive H I and electron excitation in the Cold Neutral Medium (CNM), characterized by $n = 30$ cm$^{-3}$ and $T = 100$ K, we can derive the $C^+$ cooling rate per H-atom. We can safely assume that $n \ll n_{cr}$ in the CNM, so

$$\frac{4\pi I(C^+)}{N_H} = Ah\nu \frac{g_2}{g_1} e^{h\nu/kT} \frac{[C^+]}{[H]} \sum_i \left(\frac{n}{n_{cr}}\right)_i. \qquad (10)$$

Using the definitions above relating the critical density with the cooling functions, we rewrite the $C^+$ cooling rate as

$$\frac{4\pi I(C^+)}{N_H} = \frac{[C^+]}{[H]} \sum_i (n_i L_i) = \frac{[C^+]}{[H]} \frac{(L_H + x_e L_e)}{T} p, \qquad (11)$$

where $x_e \equiv n_e/n_H$ is the fractional electron density and the pressure, and $p \equiv nT$ is in the usual astrophysical units of cm$^{-3}$ K. The temperature dependence of $L/T$ is weak, making $C^+$ an excellent probe of the interstellar pressure in the CNM. The electron density results from both the photoionization of carbon and cosmic ray ionization of hydrogen,

$$n_e = n_{C^+} + \frac{n_{C^+}}{2}\left[\left(1 + \frac{4\zeta_{CR}}{\alpha n_{C^+}^2}\right)^{1/2} - 1\right] \qquad (12)$$

The intensity ratio is, $I(C^+\ 158\ \mu m)/I(N^+\ 205\ \mu m) = 19.6 \pm 0.2$, at $|b| > 15°$. Note that the predicted level for the CNM agrees well with the observations. The calculation can be repeated for the Warm Neutral Medium (WNM), characterized by $n \sim 0.4$ cm$^{-3}$ and $T = 7000$ K. In this case the resulting intensity of $C^+$ emission is only a few percent of the CNM emission. This is not surprising since both the electron density and ion density are smaller, making collisional excitation much less effective.

Finally, we consider the contribution of the Warm Ionized Medium (WIM) to the observed 158 $\mu$m $C^+$ intensity. This was computed by Reynolds (1992) to be $I_{WIM}(C^+\ 158\mu m) = 3.6 \times 10^{-7}$ csc $|b|$ erg cm$^{-2}$ s$^{-1}$ sr$^{-1}$. Bock et al. estimate that the WIM contribution to their particular sky region is $(9 \pm 3.5) \times 10^{-8}$ erg s$^{-1}$ cm$^{-2}$ sr$^{-1}$. We derive an upper limit using an extrapolation to N(H I)=0 of the fit to the data in Figure 6, $I_{WIM}(C^+ \, 158\mu m) < 6 \times 10^{-7}$ erg s$^{-1}$ cm$^{-2}$ sr$^{-1}$ (95% CL). We conclude that the high galactic latitude C$^+$ emission predominantly arises from the CNM.

We have examined a grid of model parameters to estimate the properties of the high latitude gas cooled by the C$^+$ emission. There are many ranges of the parameters that fit the data well. For example, $T = 100$ K, $n = 10$ cm$^{-3}$, $\frac{[C^+]}{[H]} = 3.8 \times 10^{-4}$, and $\zeta_{CR} = 1 \times 10^{-17}$ s$^{-1}$ corresponds to a pressure of 1000 cm$^{-3}$ K. $T = 100$ K, $n = 20$ cm$^{-3}$, $\frac{[C^+]}{[H]} = 2.0 \times 10^{-4}$, and $\zeta_{CR} = 5 \times 10^{-18}$ s$^{-1}$ corresponds to a pressure of 2000 cm$^{-3}$ K. $T = 80$ K, $n = 15$ cm$^{-3}$, $\frac{[C^+]}{[H]} = 2.7 \times 10^{-4}$, and $\zeta_{CR} = 1 \times 10^{-16}$ s$^{-1}$ corresponds to a pressure of 1200 cm$^{-3}$ K.

When $N_H > 6.5 \times 10^{21}$ cm$^{-2}$ the slope in Figure 6 dramatically increases. Assuming the C$^+$ cooling rate remains at $(2.65 \pm 0.15) \times 10^{-26}$ erg s$^{-1}$ H atom$^{-1}$, and using the estimated total C$^+$ luminosity from our Galaxy from *COBE* of $10^{7.7}$ solar luminosities (Wright et al. 1991), we derive that the 158 $\mu$m line from C$^+$ is cooling $> 10^9$ solar masses of gas.

In Figure 7 we plot the C$^+$ line-to-continuum ratio as a function of galactic latitude. This ratio is defined using the instrument resolution bandwidth for the continuum flux. The line-to-continuum intensity ratio falls from about unity on the plane, to about one half at $|b| \sim 50°$, to the point where C$^+$ is undetectable ($< 10\%$ of the plane value) to the *COBE* FIRAS instrument above $|b| \sim 65°$. We interpret these data as indicating that the volume filling fraction of excited surfaces of cold neutral gas that preferentially give rise to C$^+$ emission decreases with increasing galactic latitude. If this is the case then the high galactic latitude far infrared emission that remains must be coming almost entirely from warmer gas. The higher excitation conditions in the warmer gas allow many atomic levels to become populated, and thus several higher energy cooling transitions can dominate over the ground state transition of C$^+$. Also, the warm gas is much less dense than the colder medium, and thus the efficiency of 158 $\mu$m cooling is greatly reduced. It is difficult to determine what fraction of high latitude far infrared emission comes from warm neutral versus warm ionized gas. This question was addressed by Boulanger & Perault (1988). They correlated H I emission with a 100 $\mu$m *IRAS* map and found a very high degree of correlation, and concluded that the far IR is coming from neutral regions. However, they also point out that their result leaves no room for the expected IR emission from the Warm Ionized Medium. This can be reconciled, however, if the H I and ionized gas are at least partially correlated with one another. Results of a survey using the *Wisconsin H$\alpha$ Mapper (WHAM)*, now under development, will help to resolve these questions.

### 4.4. Total Ionizing Photon Rate in the Galaxy

The total ionizing photon rate in the Galaxy, $\phi$, can be estimated from the total luminosity of the 205 $\mu$m N$^+$ observations from the *COBE* FIRAS, since the Galaxy is transparent at this wavelength. We revise the estimated luminosity of the 205 $\mu$m N$^+$ observations from Wright et al. (1991) upward by 30% to account for the fact that that the N$^+$ intensity follows the 1.5 rather than 1.0 power of the continuum, as was assumed in that work, now giving $L_{205\,\mu\text{m NII}} = 2.6 \times 10^{40}$ erg s$^{-1}$ inside of the solar circle. The ionizing photon rate is given by

$$\phi = \frac{L_{205\,\mu\text{m NII}}}{h\nu_{205\,\mu\text{m NII}}} \frac{[H^+]}{[N^+]} \frac{\alpha_{H^+}}{\alpha_{205\,\mu\text{m NII}}} \quad (13)$$

where $h\nu_{205\,\mu\text{m NII}} = 9.7 \times 10^{-15}$ erg, $[N^+]/[H^+] = 6.8 \times 10^{-5}$, and $\alpha_{205\,\mu\text{m NII}} = 6.7 \times 10^{-8}(T/8000\text{ K})^{-0.5}n_e$ cm$^3$ s$^{-1}$ and $\alpha_{H^+} = 6 \times 10^{-13}(T/8000\text{ K})^{-0.5}n_e$ cm$^3$ s$^{-1}$ are the recombination rates. The result is $\phi = 3.5 \times 10^{53}$ ionizing photons per second with $\sim$50% uncertainty. This is in remarkable agreement with Mezger (1978), who arrived at $3.0 \times 10^{53}$ s$^{-1}$ with 20-30% uncertainty. Our derived rate can be thought of in terms of the ionization provided by 25,000 O-stars throughout the Galaxy, each with an ionizing photon rate of $1.4 \times 10^{49}$ s$^{-1}$.

## 5. Conclusions

1. The *COBE* FIRAS has conducted an unbiased survey of the far-infrared emission from our Galaxy. The first results of this survey were reported by Wright et al. (1991). We report here the morphological distribution along the galactic plane ($b = 0°$) of the spectral line emission for all of the detected spectral lines.

2. We find that the strongest spectral line is the 158 $\mu$m fine structure transition from C$^+$. This emission peaks in the molecular ring and can be seen to follow the spiral arm structure of the Galaxy. The intensity of the C$^+$ line scales linearly with the far-infrared continuum in the plane of the Galaxy.

3. The 205 $\mu$m line of N$^+$ generally follows the 158 $\mu$m C$^+$ line distributions, but the intensities scale as $I(N^+\,205\,\mu\text{m}) \propto I(C^+\,158\,\mu\text{m})^{1.5}$ in the inner galactic plane. It has been suggested that the 1.5 power law may indicate a volume to surface area morphological sampling.

4. We present the cosecant dependence of the 158 $\mu$m fine structure transition from C$^+$ as

$$I(\text{C}^+\,158\,\mu\text{m}) = (1.43 \pm 0.12) \times 10^{-6} \csc|b| \text{ erg cm}^{-2} \text{ s}^{-1} \text{ sr}^{-1}.$$

We estimate the C$^+$ cooling rate to be $(2.65 \pm 0.15) \times 10^{-26}$ erg s$^{-1}$ H atom$^{-1}$, and conclude that the emission is arising almost entirely from the Cold Neutral Medium with a pressure of about 1000-2000 cm$^{-3}$ K. We find that the C$^+$ emission falls off much faster than the far IR continuum with increasing galactic latitude, indicating that the high galactic latitude far IR arises from a warm medium.

5. We present the cosecant dependence of the 205 $\mu$m fine structure transition from $N^+$ as

$$I(\text{N}^+\ 205\ \mu\text{m}) = (4 \pm 1) \times 10^{-8}\,\text{csc}\,|b|\ \text{erg cm}^{-2}\ \text{s}^{-1}\ \text{sr}^{-1}$$

and conclude that the emission is arising entirely from the Warm Ionized Medium.

6. We estimate the total ionizing photon rate in the inner Galaxy to be $\phi = 3.5 \times 10^{53}$ ionizing photons per second, with ~50% uncertainty, based on the 205 $\mu$m N$^+$ transition. This is in remarkable agreement with Mezger (1978), who arrived at $3.0 \times 10^{53}$ s$^{-1}$ with 20-30% uncertainty.

We are grateful to Eli Dwek, Mike Hauser, Carl Heiles, Chris McKee, Bill Reach, Ron Reynolds, and Sam Petuchowski for useful discussions. *COBE* is supported by NASA's Astrophysics Division. The Goddard Space Flight Center (GSFC), under the scientific guidance of the *COBE* Science Working Group, is responsible for the development and operation of *COBE*. We are grateful for the contributions of the entire *COBE* Science Team and the support personnel at Goddard's Cosmology Data Analysis Center (CDAC).

# FIGURE CAPTIONS

**Figure 1**:(Color Plate) The maps are projections of the full sky in galactic coordinates. The plane of the Milky Way is horizontal in the middle of the map with the galactic center at the center. Galactic longitude $\ell = 90°$ is left of center. The maps are smoothed to 10° resolution. *(top)*: A full sky map of 158 $\mu$m $C^+$ emission from the *COBE* FIRAS experiment. *(bottom)*: A full sky map of 205 $\mu$m $N^+$ emission from the *COBE* FIRAS experiment.

**Figure 2**: Longitude profiles of the intensity solution for spectral lines at given wavelengths for the entire galactic plane. The intensity in each bin represents an average over 5° in galactic longitude and ±5° in galactic latitude. Error bars include detector noise and gain uncertainties, which affect the accuracy of the continuum removal. There are very few measurements in the $\ell = 90°$, $b = 0°$ bin, so we caution against overinterpretation of results based only on this data point. (a) The detected $C^+$ $2p$:$^2P_{3/2} - ^2P_{1/2}$ fine structure ground state transition at 158 $\mu$m. The dashed line is the far infrared continuum emission at that wavelength and within the same effective bandwidth. (b) The undetected $O_2$ $3_2 - 1_2$ at 705.8 $\mu$m. (c) The detected $N^+$ $2p^2$ $:^3P_1 - ^3P_0$ fine structure ground state transition at 205.3 $\mu$m. The dashed line is 0.07 times the far infrared continuum emission at that wavelength and within the same effective bandwidth. (d) The detected $N^+$ $2p^2$ $:^3P_2 - ^3P_1$ transition at 121.9 $\mu$m. (e) The possibly detected (see text) $O^0$ $2p^4$ $:^3P_0 - ^3P_1$ transition at 145.5 $\mu$m. (f) The undetected $Si^0$ $3p^2$ $:^3P_1 - ^3P_0$ $Si^0$ transition at 129.7 $\mu$m. (g) The undetected ortho-$H_2O$ $1_{10} - 1_{01}$ ground state transition at 538.3 $\mu$m. (h) The possibly detected (see text) para-$H_2O$ $1_{11} - 1_{00}$ ground state transition at 269.3 $\mu$m. (i) The undetected CO J=1-0 transition. (j) The detected CO J=2-1 transition. (k) The detected CO J=3-2 transition. (l) The detected CO J=4-3 transition. (m) The detected CO J=5-4 transition. (n) The undetected CO J=6-5 transition. (o) The detected $C^0$ $2p^2$ $:^3P_1 - ^3P_0$ transition at 609.1 $\mu$m. (p) The detected $C^0$ $2p^2$ $:^3P_2 - ^3P_1$ transition at 370.4 $\mu$m.

**Figure 3**: An excellent correlation exists between the intensity of the $C^+$ and $N^+$ intensities in the Galactic plane, in the sense that $I(N^+) \propto I(C^+)^{1.5}$.

**Figure 4**: $I(C^+) \propto I(FIR)^{0.95}$ while $I(N^+) \propto I(FIR)^{1.53}$ in the Galactic plane.

**Figure 5**: (Color Plate) The maps are projections of the full sky in galactic coordinates. The plane of the Milky Way is horizontal in the middle of the map with the galactic center at the center. Galactic longitude $\ell = 90°$ is left of center. The maps are smoothed to 10° resolution. *(top)*: The full sky distribution of neutral hydrogen H I emission, described by Dickey & Lockman (1990). *(bottom)*: The $C^+$ cooling rate per hydrogen atom, $4\pi I(C^+ \ 158 \ \mu m)/N(H \ I)$.

**Figure 6**: *(top)*: The observed 158 $\mu$m $C^+$ intensity versus the H I column density, $N(H)$ cm$^{-2}$, as described by Dickey & Lockman (1990). *(bottom)*: Data restricted to at least 30° from the galactic center and $N(H) < 6 \times 10^{21}$ H-atoms cm$^{-2}$. The three lines correspond to our best estimate of the $C^+$ cooling rate of $(2.65 \pm 0.15) \times 10^{-26}$ erg s$^{-1}$ H atom$^{-1}$ and its 95% confidence upper and lower bounds. This assumes that the H I measurements are accurate and that $C^+$ and H I emission are coextensive.

**Figure 7**: The $C^+$ line-to-continuum intensity ratio as a function of galactic latitude. The far IR continuum is defined as the best fit continuum intensity within the same effective bandwidth (i.e. instrument resolution bandwidth) as the $C^+$ line emission. The $C^+$ emission decreases with increasing galactic latitude until it becomes undetectable at $|b| > 65°$.

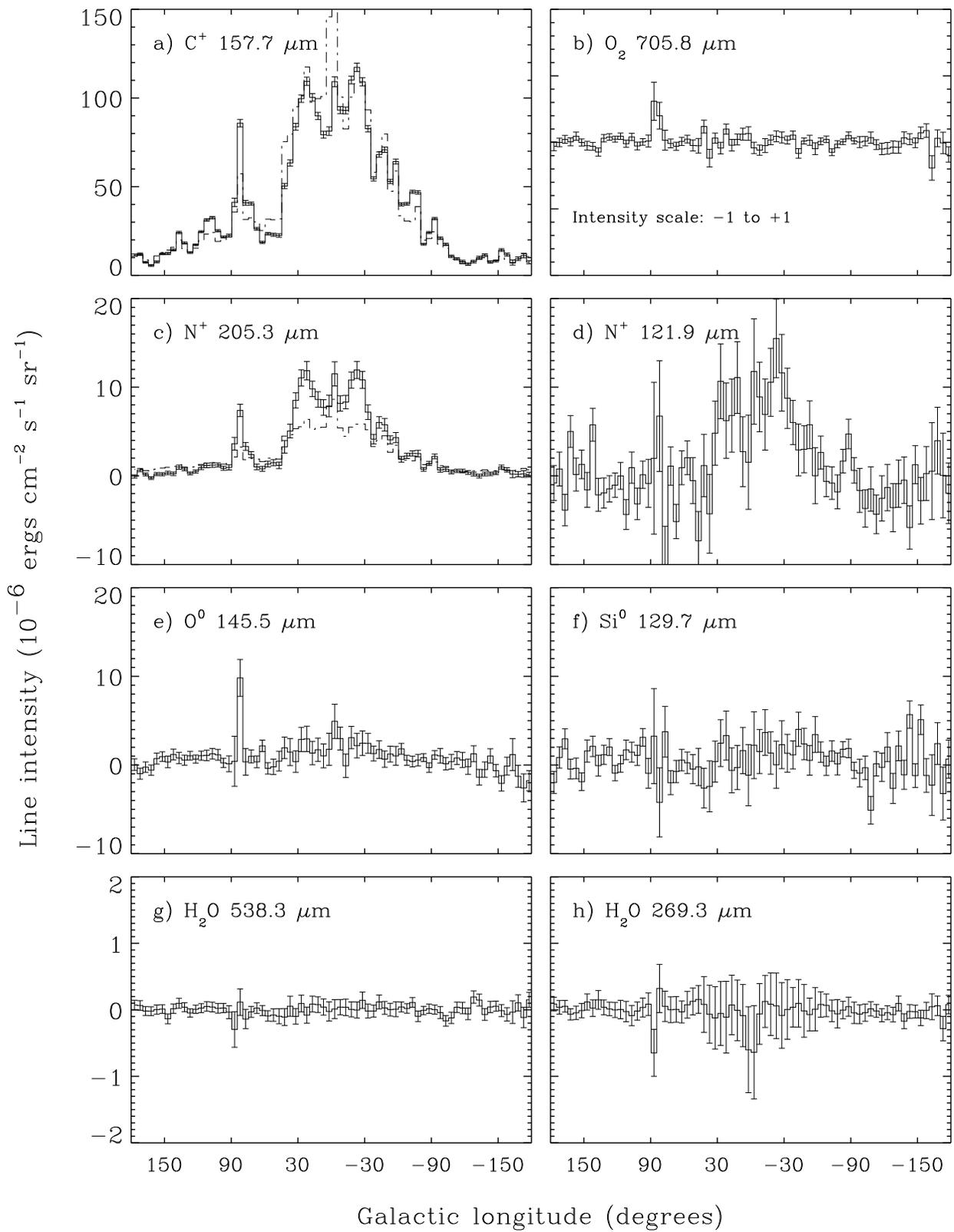

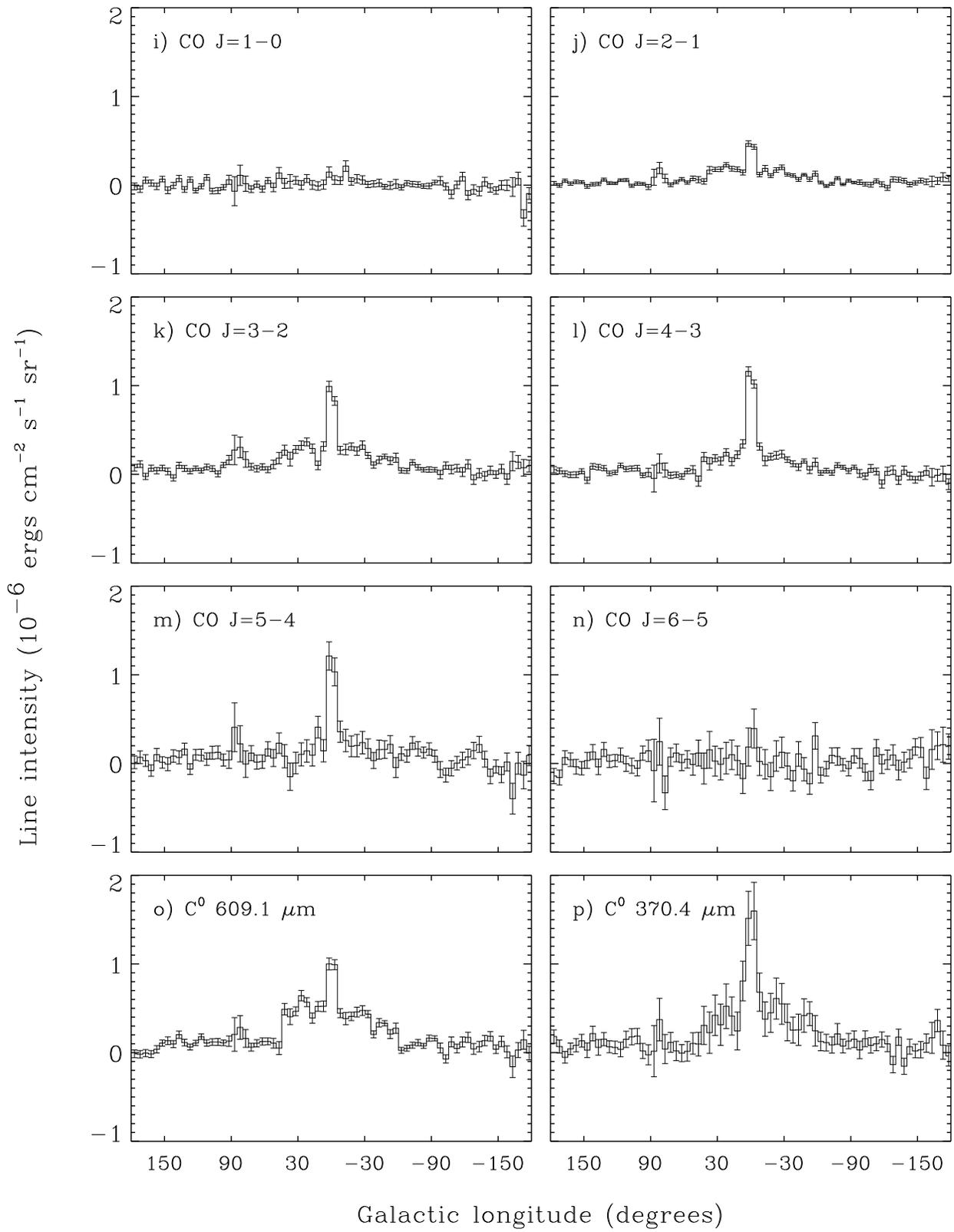

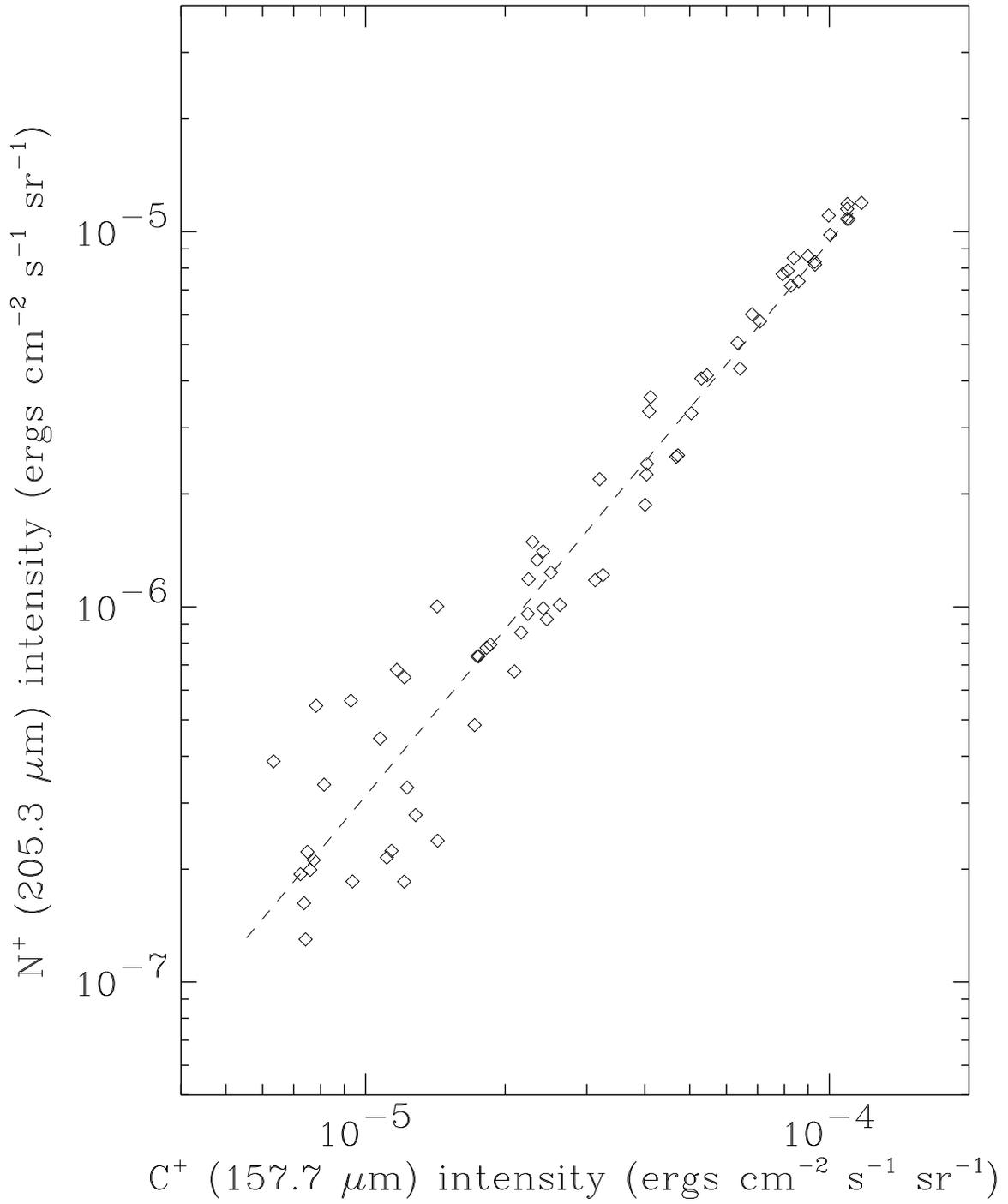

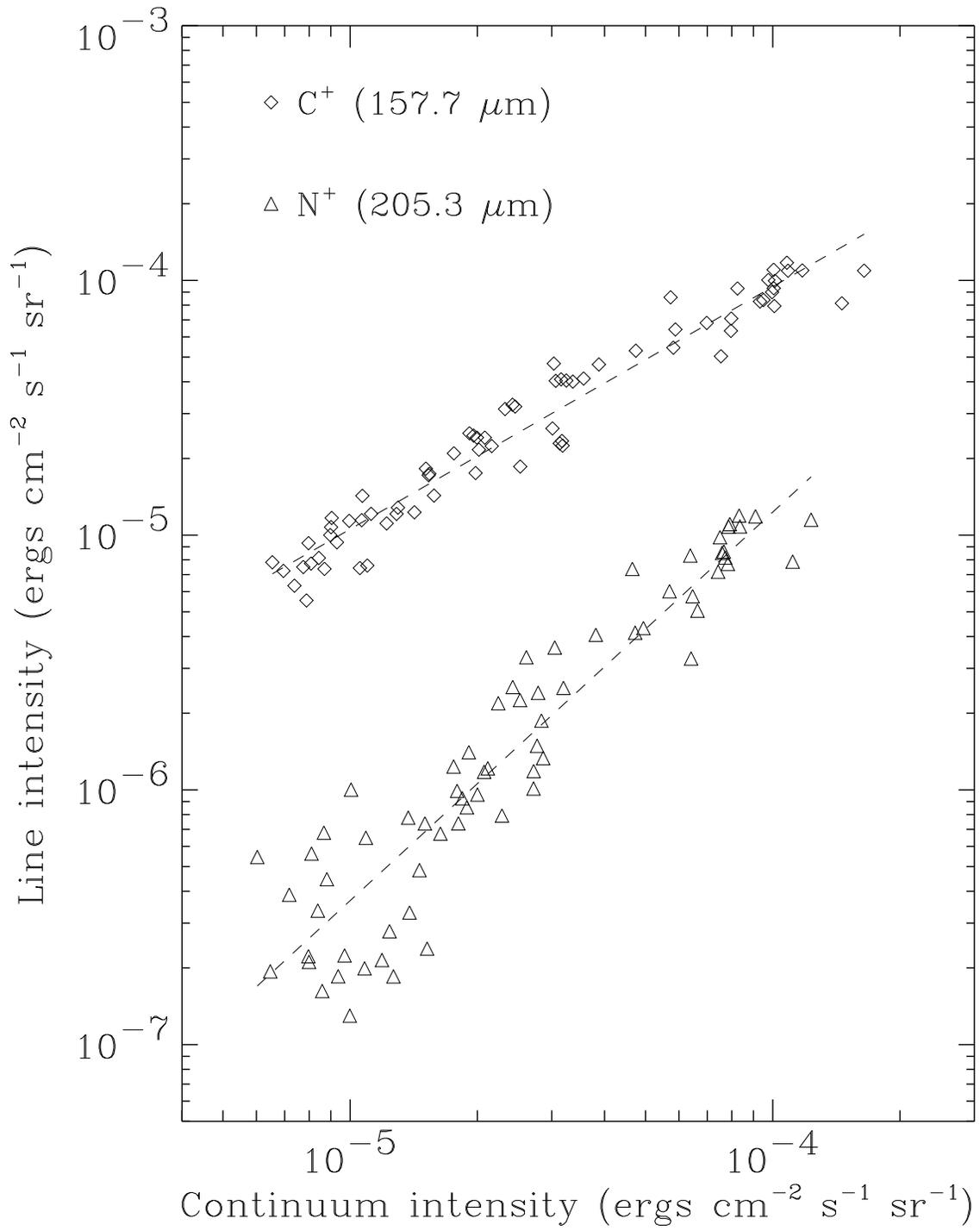

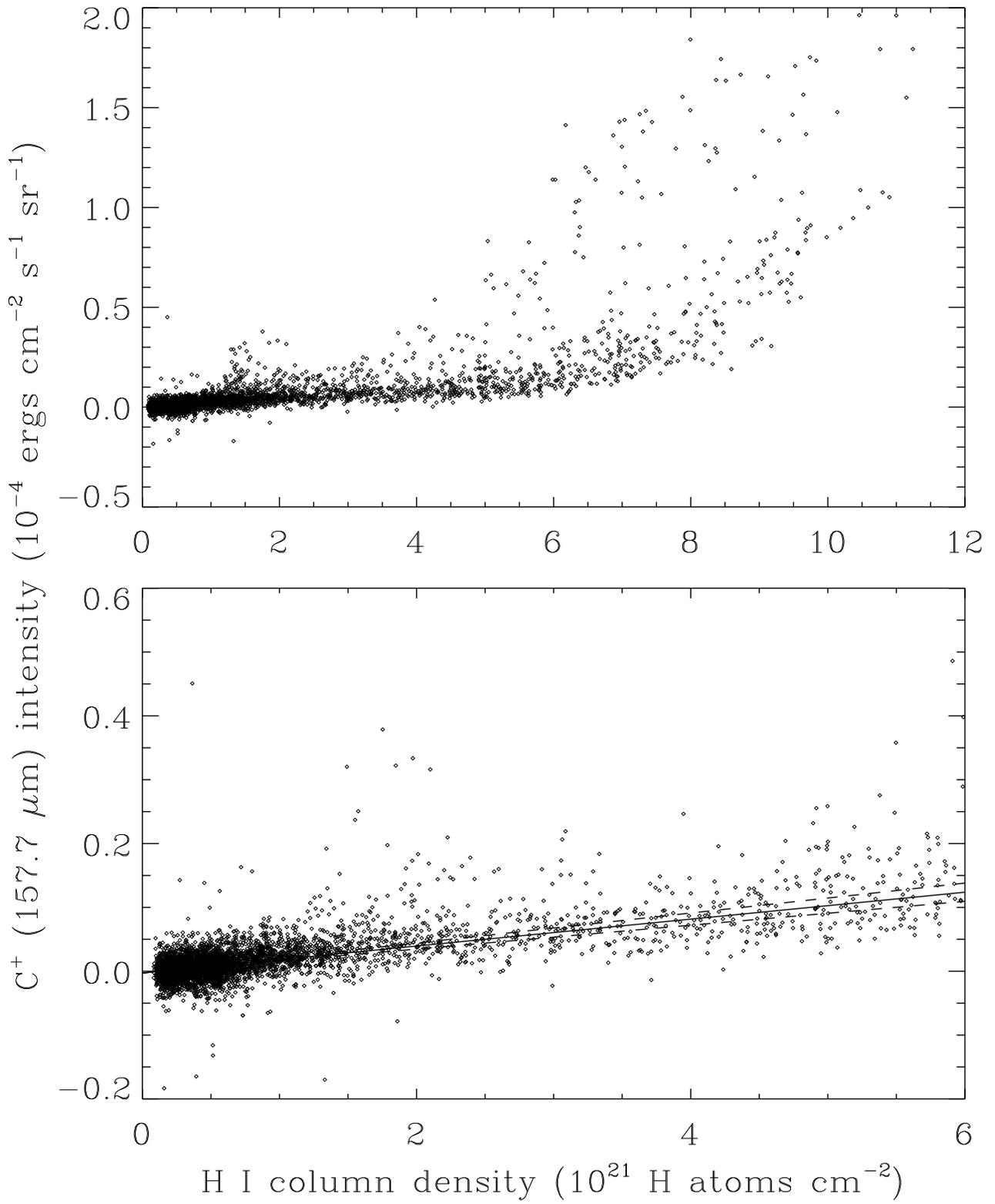

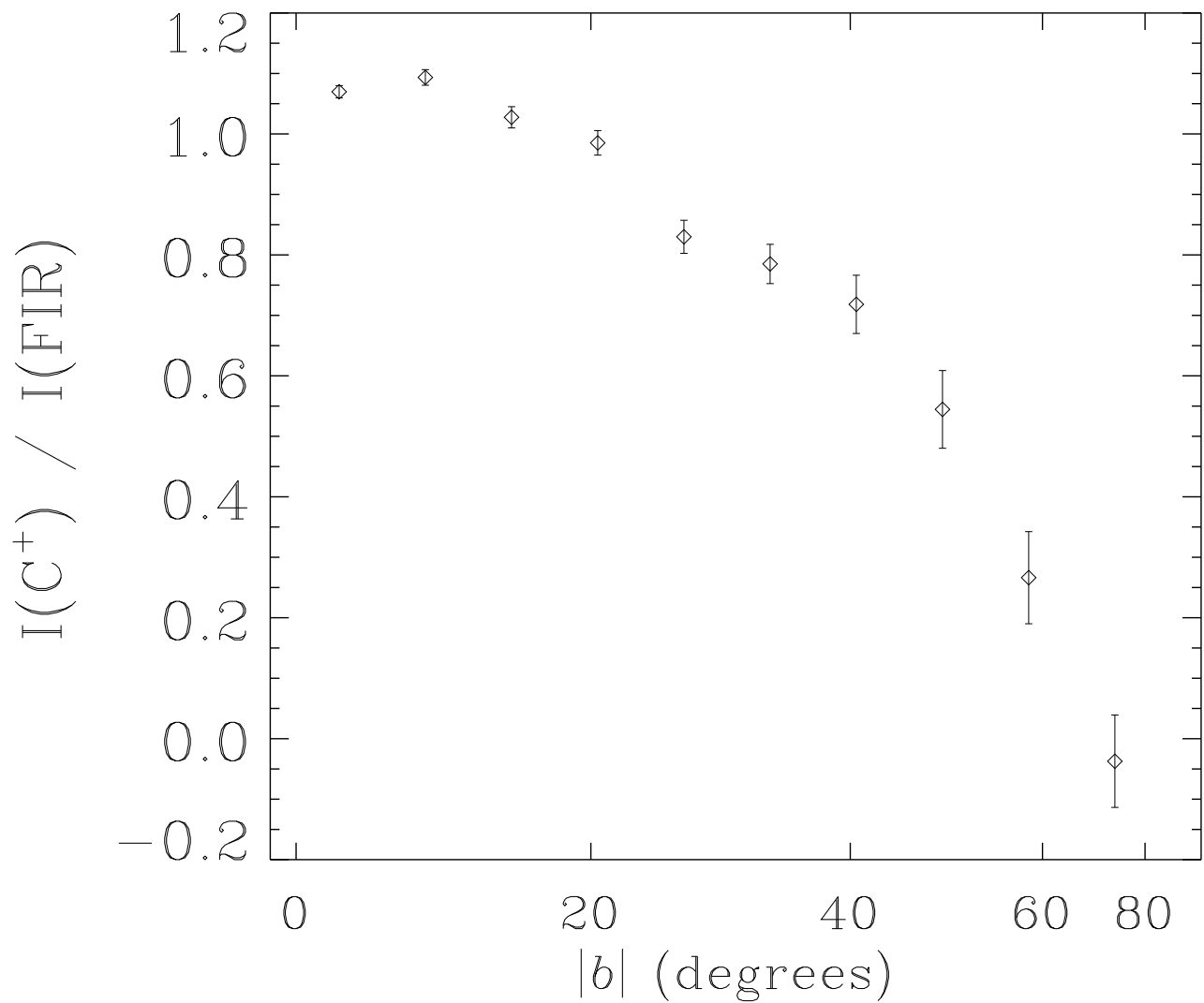